\documentclass[conference]{IEEEtran}
\IEEEoverridecommandlockouts
\usepackage{cite}
\usepackage{dblfloatfix}
\usepackage{overpic}
\usepackage{amsmath,amssymb,amsfonts}
\usepackage{graphicx}
\usepackage{textcomp}
\usepackage{xcolor}
\usepackage{float}
\usepackage{amsthm}
\usepackage{graphicx}
\usepackage{epstopdf}
\usepackage{amsmath,bm}
\usepackage{amsfonts}
\usepackage{amssymb}
\usepackage{color}
\usepackage{multirow}
\usepackage{multicol}
\usepackage{soul,xcolor}
\usepackage{algorithm}

\usepackage{algorithmic}

\usepackage{caption}
\usepackage{subcaption}
\usepackage[justification=centering]{caption}

\ifCLASSOPTIONcompsoc
  \usepackage[caption=false,font=normalsize,labelfont=sf,textfont=sf]{subfig}
\else
  \usepackage[caption=false,font=footnotesize]{subfig}
\fi

\theoremstyle{plain}

\newtheorem{remark}{Remark}

\newcounter{relctr} 
\everydisplay\expandafter{\the\everydisplay\setcounter{relctr}{0}} 

\def\Htran{\mbox{\tiny $\mathrm{H}$}}
\def\Ttran{\mbox{\tiny $\mathrm{T}$}}
\def\CN{\mathcal{N}_{\mathbb{C}}}

\begin{document}

\title{DoA Estimation using MUSIC with Range/Doppler Multiplexing for MIMO-OFDM Radar 
}


\author{\IEEEauthorblockN{Murat Babek Salman, Emil Björnson}
\IEEEauthorblockA{\textit{School of Electrical Engineering and Computer Science, KTH Royal Institute of Technology, Kista, Sweden}}
\IEEEauthorblockA{
mbsalman@kth.se, emilbjo@kth.se}


}

\maketitle
\begin{abstract}

Sensing emerges as a critical challenge in 6G networks, which require simultaneous communication and target sensing capabilities. State-of-the-art super-resolution techniques for the direction of arrival (DoA) estimation encounter significant performance limitations when the number of targets exceeds antenna array dimensions. This paper introduces a novel sensing parameter estimation algorithm for orthogonal frequency-division multiplexing (OFDM) multiple-input multiple-output (MIMO) radar systems. The proposed approach implements a strategic two-stage methodology: first, discriminating targets through delay and Doppler domain filtering to reduce the number of effective targets for super-resolution DoA estimation, and second, introducing a fusion technique to mitigate sidelobe interferences. The algorithm enables robust DoA estimation, particularly in high-density target environments with limited-size antenna arrays. Numerical simulations validate the superior performance of the proposed method compared to conventional DoA estimation approaches.
\end{abstract}

\section{Introduction}


The emergence of 6G network technology has brought integrated sensing and communication (ISAC) to the forefront of wireless research. ISAC systems address the growing need for enhanced environmental awareness and support novel use cases in future networks \cite{10217169}. One key benefit is improved communication security, as ISAC can detect potential eavesdroppers, making transmissions more robust against threats \cite{10373185}. In the expanding Internet of Things (IoT), ISAC enhances sensing accuracy, which is crucial for applications like asset tracking and smart city infrastructure \cite{9737357}.

The strength of ISAC lies in its efficient use of shared hardware, spectrum, and time resources, reducing complexity and costs in network deployments \cite{10214237}. By adding sensing capabilities to a network originally built to provide wide-area coverage for communication services, sensing services also become widely available. Moreover, by integrating sensing and communication, ISAC optimizes spectrum utilization, which is vital given the increasing scarcity of available spectrum. A critical component in 6G is the integration of multiple-input multiple-output (MIMO) technology, which enhances performance through spatial diversity and resolution \cite{10124714}. MIMO-enabled ISAC systems are highly flexible, adaptable, and suitable for the diverse requirements of 6G networks. 


The development of effective ISAC systems requires a waveform that simultaneously supports both functions without compromising performance in either domain, posing significant design challenges \cite{WaveformDesign}. Orthogonal Frequency Division Multiplexing (OFDM) has proven to be effective for communication systems in rich scattering environments. Recently, the integration of the MIMO-OFDM concept within the ISAC framework, aiming to leverage its usefulness in communication to address the complex requirements of these dual-purpose systems, has been investigated. Also, several parameter estimation techniques have been proposed for MIMO-OFDM systems in the literature \cite{9838368,MusaFurkan,10594515,10036975,10634583}. In \cite{9838368}, \cite{MusaFurkan} and \cite{10594515}, the MUltiple SIgnal Classification (MUSIC) algorithm is used for the first stage to estimate the direction-of-arrival (DoA) of the targets. To estimate the delay and the velocity of the targets, the two-dimensional Discrete Fourier Transform (2D-DFT) method is utilized in \cite{9838368}. In \cite{MusaFurkan}, a maximum likelihood estimator following receive beamforming is developed. In \cite{10594515}, a two-stage MUSIC algorithm is proposed to reduce the computational complexity of the DoA estimation. However, these methods only perform well if the number of targets is less than the number of antennas. To increase the number of simultaneously localized targets, DFT-based angle estimation methods have been proposed in \cite{10036975} and \cite{10634583}. In these works, DFT-based DoA estimation is performed to detect more targets. In \cite{10036975}, coarse angle estimation is carried out by using DFT along the antenna dimension; then, estimation refinement is performed via sparse signal recovery with a DFT basis vector. The performance of this method is further improved in \cite{10634583} by performing joint angle-delay-Doppler estimation by using a similar DFT approach. However, the performance of DFT-based methods suffers from low angular resolution.

To improve the DoA estimation accuracy, high-resolution angle estimation is necessary. Hence, the use of super-resolution algorithms such as MUSIC is inevitable. This paper introduces a novel algorithm that first discriminates targets in either the delay or Doppler domains before performing DoA estimation when the number of antennas at the access points (APs) in a small cell or distributed network is limited. This preprocessing significantly reduces the number of targets processed simultaneously in the angular domain, enabling efficient MUSIC implementation. Furthermore, we introduce a fusion methodology that effectively mitigates ambiguities caused by sidelobe interference in the delay/Doppler processing stage, enhancing the overall robustness of target sensing. Improved DoA estimation performance thanks to the proposed method is verified by the numerical simulations.

\section{Transmit Signal and Radar Echo Models}

In this section, we present the system model, which focuses on a single AP within a distributed network. Such a setup is common in modern wireless architectures, including small cell deployments and distributed MIMO systems, where multiple APs collaborate to serve users across the coverage area. The AP under consideration is equipped with $N_{\rm t}$ transmit and $N_{\rm r}$ receive antennas. Our objective is to localize $I$ targets by estimating their delay, Doppler shift, and DoA. Since the AP is operated in joint communication and sensing mode, an OFDM waveform is transmitted to perform both tasks. The transmitter sends $M$ OFDM symbols, which can be expressed in the complex baseband as a continuous-time signal
\begin{equation}
    {\bf x}(t) = \frac{1}{\sqrt{N_c}} \sum_{m=0}^{M-1} \sum_{n = 0}^{N_c-1} {\bf x}_{mn} e^{j2\pi \Delta f nt} \operatorname{rect}\left(\frac{t-mT_{\rm sym} }{T_{\rm sym}}\right),
\end{equation}
where ${\bf x}_{mn} \in \mathbb{C}^{N_{\rm t} \times 1}$ is the data vector at the $m^{\rm th}$ symbol and $n^{\rm th}$ subcarrier, ${N_c}$ is the number of subcarriers, $\Delta f$ is the subcarrier spacing, $T_{\rm sym}$ is the symbol duration, which can be defined as $T_{\rm sym}\triangleq \frac{1}{\Delta f} + \frac{N_{cp}}{\Delta f N_c}$ where $N_{cp}$ is the length of cyclic prefix. The total bandwidth of the waveform is $B = \Delta f N_c$, and $\operatorname{rect}(\cdot)$ is the rectangular function. The up-converted radio frequency (RF) signal with carrier frequency $f_c$ is expressed as 
\begin{equation}
    {\bf x}_{\rm RF} (t) = {\bf x}(t) e^{j 2 \pi f_c t}.
\end{equation}
\subsection{Radar Echo Signal}
Next, the model for the received radar echo signal is presented, which will later be utilized for parameter estimation. The received echo signal from $I$ moving targets with DoAs $\{\theta_i \}_{i=1}^I$ can be expressed as
\begin{equation}
\begin{split}
 &{\bf y}_{\rm echo}(t)\\
 &= \sum_{i=1}^I \alpha_i {\bf a}_{\rm R} (\theta_i) {\bf a}^{\Ttran}_{\rm T} (\theta_i) {\bf x}\left(t-\tau_i(t)\right) e^{j2\pi f_c \left(t-\tau_i(t)\right)} + {\boldsymbol \eta}(t),
\end{split}
\end{equation}
where $\alpha_i$ combines the effects of the two-way path-loss and radar cross-section of the target, $\tau_i(t) = \tau_i - \nu_i t$, $\tau_i = 2R_i/c$ is the time delay containing the Doppler shift $\nu_i = 2v_i/c $, and $v_i$ is the radial speed of the target. Finally, ${\boldsymbol \eta}(t)$ is the additive white Gaussian noise (AWGN) with constant power spectral density $N_o/2$. After analog matched filtering and down-conversion, the received signal for the $m^{\rm th}$ OFDM symbol after sampling at time $t= mT_{\rm sym} + \frac{p}{\Delta f N_c}$ based on the narrowband assumption and ignoring the constant terms stated in \cite{MusaFurkan2} can be written as
\begin{equation}
\begin{split}
    {\bf y}_{mp} = &\frac{1}{\sqrt{N_c}}\sum_{i=1}^I \alpha_i {\bf a}_{\rm R} (\theta_i) {\bf a}^{\Ttran}_{\rm T} (\theta_i)  \\
    &\sum_{n=0}^{N_c-1} {\bf x}_{mn} e^{j 2\pi \frac{np}{N_c}} e^{-j 2\pi \Delta f \tau_i n} e^{j 2\pi f_i^d m T_{\rm sym}} + {\boldsymbol \eta}_{mp},
\end{split}
\end{equation}
where $f_i^d = f_c \frac{2v_i}{c}$ is the Doppler shift and $ {\boldsymbol \eta}_{mp}$ is the sampled noise with distribution ${\boldsymbol \eta}_{mp} \sim \CN (0, BN_o {\bf I}_{N_r})$. To estimate both the range and velocity using ${\bf y}_{mp}$, the standard DFT-based method is used, where the signal in the DFT domain for the $n^{\rm th}$ bin can be expressed as \cite{9838368,10634583}
\begin{equation}
\begin{split}
    &\bar{{\bf y}}_{mn}\\
    &= \sum_{i=1}^I \alpha_i {\bf a}_{\rm R} (\theta_i) {\bf a}^{\Ttran}_{\rm T} (\theta_i)  {\bf x}_{mn} e^{-j2\pi \Delta f \tau_i n} e^{j2\pi f_i^d m  T_{\rm sym}} + \bar{{\boldsymbol \eta}}_{mn}. \label{eq:rxDFT}
\end{split}
\end{equation}
It can be seen from \eqref{eq:rxDFT} that the received signal $\bar{{\bf y}}_{mn}$ can be expressed as a superposition of DFT basis functions for both range with component $e^{-j2\pi \Delta f \tau_i n}$ over subcarriers and Doppler shift with component $e^{-j2\pi f_i^d m  T_{\rm sym}}$ over symbols. In the following sections, we present a dual-stage localization methodology: first estimating delay-Doppler parameters using 2D-DFT method, then applying selective interference filtering for DoA determination from the observation $\bar{{\bf y}}_{mn}$. This framework enables robust multi-target detection with improved signal-to-interference-plus-noise ratio.
\section{Range/Doppler estimation}
In this section, the delay/Doppler processing scheme for range/velocity estimation is derived without prior DoA information. To jointly estimate the range and Doppler parameters from $\bar{{\bf y}}_{mn}$, we employ the 2D-DFT  technique. Firstly, matched filtering (MF) is applied to the received signal $\bar{{\bf y}}_{mn}$ as
\begin{equation}
\begin{split}
   &\bar{{\bf Y}}_{mn} = \sum_{i=1}^I \alpha_i {\bf a}_{\rm R} (\theta_i) {\bf a}^{\Ttran}_{\rm T} (\theta_i) e^{-j2\pi \Delta f \tau_i n} e^{j2\pi f_i^d m  T_{\rm sym}} {\bf x}_{mn} {\bf x}_{mn}^{\Htran} \\
   & \quad  + \bar{{\boldsymbol \eta}}_{mn} {\bf x}_{mn}^{\Htran},
\end{split}
\end{equation}
where $\bar{{\bf Y}}_{mn}$ represents the $N_{\rm r} \times N_{\rm t}$ spatial antenna channels. We can express each virtual channel ($(k,l)^{\rm th}$ element of $\bar{{\bf Y}}_{mn}$) as
\begin{equation}
\begin{split}
    &\bar{{\bf Y}}_{mn}[k,l]\\
    &= \sum_{i=1}^{I} \alpha_i { a}_{{\rm R},k} (\theta_i) \left[ \sum_{p=1}^{N_{\rm t}} { a}_{{\rm T},p} (\theta_i) {\bf x}_{mn}[p] {\bf x}_{mn}^{*}[l] \right]  w_i^n \omega_{d^i}^m \\
    &\quad + \bar{{\boldsymbol \eta}}_{mn}[k] {\bf x}_{mn}^{*}[l], \label{eq:RadarChannels}
\end{split}
\end{equation}
where $ w_i \triangleq  e^{-j2\pi \Delta f \tau_i}$ and $\omega_{d^i} \triangleq e^{j2\pi f_i^d T_{\rm sym}}$. Note that when the antenna index $p$ aligns with the true index $l$, the data-dependent term $|{\bf x}_{mn}[l]|^2$ within the summation becomes positive, facilitating a coherent combination of signal components. This coherent integration yields a distinctive peak in the range-Doppler plane following the 2D-DFT. For each virtual channel branch, we can form the $M\times N_c$ data matrix as
\begin{equation}
\begin{split}
    \bar{{\bf Y}}[k,l] = & \sum_{i = 1}^I  \alpha_i { a}_{{\rm R},k} (\theta_i)   \sum_{p=1}^{N_{\rm t}}  { a}_{{\rm T},l} (\theta_i) {\boldsymbol \Omega}_i ({\bf X}[p] \odot {\bf X}^*[l])  {\bf W}_i\\
    &+ {\boldsymbol \zeta[k]} \odot {\bf X}^*[l], \label{eq:DataMatrix}
\end{split}
\end{equation}
where $\odot$ is the Hadamard product, $\left[{\bf X}[l]\right]_{mn} = {\bf x}_{mn}[l]$ and we define the diagonal matrices $[{\boldsymbol \Omega}_i]_{mm} \triangleq \omega_{d^i}^m$ and $[{\bf W}_i]_{nn} \triangleq w_i^n$. Due to the lack of DoA knowledge at the radar receiver, beamforming before detection is not possible. Hence, non-coherent processing through different antenna channels will be used to perform the detection. The range-Doppler map for each virtual branch is extracted individually and combined non-coherently. The range-Doppler for $(k,l)^{\rm th}$ channel is extracted via the 2D-DFT as
\begin{equation}
    {\boldsymbol \chi }[k,l] = {\bf F}_M^{\Ttran} \bar{{\bf Y}}[k,l] {\bf F}_{N_c}, \label{eq:chiKL}
\end{equation}
where ${\bf F}_M \in \mathbb{C}^{M \times M}$ and ${\bf F}_{N_c} \in \mathbb{C}^{N_c \times N_c}$ are DFT matrices with elements $[{\bf F}_M]_{mn} = e^{j2\pi \frac{mn}{M}}$ and $[{\bf F}_{N_c}]_{mn} = e^{j2\pi \frac{mn}{N_c}} $.

The detection of the range and velocities of targets can be performed via a non-coherent combination of the delay-Doppler maps of each virtual channel branch \cite{8726131}. The non-coherent combination can be expressed as
\begin{equation}
    \bar{{\boldsymbol \chi }} = \sum_{k=1}^{N_{\rm r}} \sum_{l=1}^{N_{\rm t}} |{\boldsymbol \chi }[k,l]|^2. \label{eq:NCI}
\end{equation}
Following the non-coherent combination, a peak detection algorithm is applied to $\bar{{\boldsymbol \chi }}$ to identify potential targets. Subsequently, hypothesis testing is performed on these detected peaks to determine the presence or absence of targets and to estimate their range and velocity parameters.

\begin{remark}
    For multi-target scenarios, it is probable that some targets are not resolvable in the delay or Doppler domains. However, it can be assumed that they can be resolved in at least one of these domains. By applying selective delay and Doppler multiplexing, the proposed method effectively reduces the number of active targets processed simultaneously. This strategy enables independent DoA estimation in each domain, ultimately allowing the detection of more targets than traditional antenna array size limitations would permit. The proposed technique suppresses inter-target interference, enhancing estimation accuracy in complex signal environments.
\end{remark}

\section{DoA Estimation via Range/Doppler Filtering}

In this section, details of the proposed DoA estimation after range/Doppler processing are introduced. For this purpose, the received signal is processed independently in the delay and Doppler domains. Then, the MUSIC algorithm is employed to detect the DoA corresponding to each delay and Doppler estimate extracted via \eqref{eq:NCI}. Then, the obtained DoAs corresponding to each target are combined so that a more accurate estimation is performed. As the first step, we construct the received signal vector for each transmit channel as
\begin{equation}
\begin{split}
    &\tilde{\bf y}_{mn,l} = \bar{\bf Y}_{mn} {\bf e}_{l}\\
    &= \sum_{i=1}^I \alpha_{i} {\bf a}_{\rm R} (\theta_{i}) \left( {\bf a}^{\Ttran}_{\rm T} (\theta_{i})  {\bf x}_{mn} {\bf x}_{mn}^{*}[l]\right) w_i^n \omega_{d^i}^m +  \bar{{\boldsymbol \eta}}_{mn}  {\bf x}_{mn}^{*}[l], \label{eq:TxCombining}
\end{split}
\end{equation}
where ${\bf e}_{l}$ is the elementary vector whose elements are $0$ except for $l^{\rm th}$ element, which is $1$. From \eqref{eq:TxCombining}, it can be observed that phase difference across the elements of the vector $\tilde{\bf y}_{mn}$ stems from only the spatial variation. Signal components from multiple targets contribute to the received signal, which yields inter-target interference, respectively; hence, through selective delay and Doppler filtering, we propose to isolate individual targets for the accurate direction of arrival estimation. 
\vspace{-5mm}
\subsection{Delay Multiplexing}
To eliminate the interference due to targets with different delays, filtering in the delay domain is proposed to be used. Doing so reduces the number of effective targets, whose DoAs are aimed to be estimated. We can extract signals resulting from this specific delay from the received signal by filtering over a single OFDM block. Let us focus on a delay peak of \eqref{eq:NCI} with index $\hat{\tau}_i$. Then, the \emph{temporal MF} can be applied to the signal after MF, which can be expressed as
\begin{equation}
   \begin{split}
       &\tilde{\bf y}_{m,l}^{\hat{\tau}_i} = \frac{1}{{\sqrt{N_c}}}  \sum_{n = 0}^{N_c-1} \tilde{\bf y}_{mn,l}  e^{j2\pi \Delta f{\hat \tau_{i}} n} \\
       &= \sum_{i'=1}^I {\bf a}_{\rm R} (\theta_{i'})   \sum_{n=0}^{N_c-1}  \tilde{\alpha}_{i',mn,l} e^{-j2\pi \Delta f (\tau_{i'}-\hat{\tau}_i) n}   \omega_{d^{i'}}^m + \tilde{{\boldsymbol \eta}}_{m,l}, \label{eq:temporalMF}
   \end{split}
\end{equation}
where $ \tilde{{\boldsymbol \eta}}_{m,l} \triangleq \frac{1}{{\sqrt{N_c}}}  \sum_{n = 0}^{N_c-1}  \bar{{\boldsymbol \eta}}_{mn} {\bf x}_{mn}^{*} {{\bf e}_{l}} e^{j2\pi \Delta f{\hat \tau_{i'}} n}$, and $\tilde{\alpha}_{i,mn,l} \triangleq \frac{\alpha_{i}}{\sqrt{N_c}} {\bf a}^{\Ttran}_{\rm T} (\theta_{i})  {\bf x}_{mn} {\bf x}_{mn}^{*} {{\bf e}_{l}}$. Note that $\tilde{\alpha}_{i,mn,l}$ can be explicitly written as $\tilde{\alpha}_{i,mn} = \frac{\alpha_{i}}{\sqrt{N_c}}\sum_{l' = 1}^{N_{\rm t}} {\bf a}_{{\rm T}, l'}(\theta_i)  {\bf x}_{mn}[l'] {\bf x}_{mn}[l]$. Assuming that the number of subcarriers $N_c$ is large enough, the term containing $|{\bf x}_{mn}[l]|^2$ becomes dominant. Hence, omitting the remaining terms, \eqref{eq:temporalMF} can be written as
\begin{equation}
    \begin{split}
        \tilde{\bf y}_{m,l}^{\hat{\tau}_i} &\approx \sum_{i'=1}^{I}  \alpha_{i}\frac{{\bf a}_{\rm R} (\theta_{i'})   }{\sqrt{N_c}} \left[ \sum_{n=0}^{N_c-1}  \tilde{\bf x}_{mn} [l]e^{-j2\pi \Delta f (\tau_{i'}-\hat{\tau}_i) n}  \right] \omega_{d^{i'}}^m\\
        & + \tilde{{\boldsymbol \eta}}_{m,l} \\
        & \overset{\textrm{(a)}}{\approx} \sum_{i' \in \mathcal{I}_{\tau_i}}  \alpha_{i}\frac{{\bf a}_{\rm R} (\theta_{i'}) }{\sqrt{N_c}} \left[ \sum_{n=0}^{N_c-1}  \tilde{\bf x}_{mn}[l] \right] \omega_{d^{i'}}^m + \tilde{{\boldsymbol \eta}}_{m,l},
        \label{eq:temporalMF2}
    \end{split}
\end{equation}
where  $\tilde{\bf x}_{mn} [l] \triangleq {\bf a}_{{\rm T},l} (\theta_{i'})|{\bf x}_{mn}[l]|^2$, and in (a), we have used the fact that only a subset of the targets are in proximity to the intended target in the delay domain such that they fall into the side-lobe. Target DoA estimation may be compromised when interfering targets have stronger radar cross section (RCS) in sidelobes. However, by implementing Doppler filtering, we can effectively eliminate interference from targets previously unresolvable in the delay domain, enabling accurate DoA determination. 
From \eqref{eq:temporalMF2}, it can be observed that $\tilde{\bf y}_{m,l}^{\hat{\tau}_i}$ has a similar form to be used to estimate the DoA using the MUSIC algorithm. Hence, the next step is constructing the sample autocorrelation matrix $\hat{\bf R}^{\hat{\tau}_i}$ to obtain the noise subspace as 
%
%
%
\begin{equation}
    \begin{split}
        & \hat{\bf R}^{\hat{\tau}_i} =  \sum_{l=0}^{N_{\rm t}-1} \sum_{m=0}^{M-1} \tilde{\bf y}_{m,l}^{\hat{\tau}_i} \left(\tilde{\bf y}_{m,l}^{\hat{\tau}_i} \right)^{\Htran}\\
        & \approx \sum_{l=0}^{N_{\rm t}-1} \sum_{m=0}^{M-1}  \sum_{i' \in \mathcal{I}_{\tau_i}} \sum_{i'' \in \mathcal{I}_{\tau_i}} \alpha_{i'} \alpha_{i''}^* \frac{{\bf a}_{\rm R} (\theta_{i'}) {\bf a}_{\rm R}^{\Htran} (\theta_{i''}) }{{N_c}} |\tilde{x}_m^l|^2 \omega_{d^{i'-i''}}^m\\
        &+ \sum_{l=0}^{N_{\rm t}-1} \sum_{m=0}^{M-1}  \tilde{{\boldsymbol \eta}}_{m,l} \tilde{{\boldsymbol \eta}}_{m,l}^{\Htran }, \label{eq:R1Delay}
    \end{split}
\end{equation}
where $\tilde{x}_m^l \triangleq  \sum_{n=0}^{N_c-1}  \tilde{\bf x}_{mn}[l] $. Since $\omega_{d^{i'-i''}}^m$ represents a complex sinusoidal term and the users in the same range are assumed to be resolvable in the Doppler domain, the summation over these oscillating terms averages to approximately zero, allowing us to neglect the cross-terms in \eqref{eq:R1Delay}. Hence, we can write the auto-correlation function as
%
\begin{equation}
    \hat{\bf R}^{\hat{\tau}_i} \approx  \sum_{i' \in \mathcal{I}_{\tau_i}}  |\alpha_{i'}|^2 {\bf a}_{\rm R} (\theta_{i'}) {\bf a}_{\rm R}^{\Htran} (\theta_{i'})  \sum_{l=0}^{N_{\rm t}-1} \sum_{m=0}^{M-1}  \frac{|\tilde{x}_m^l|^2}{{N_c}}   + {N_{\rm t}} M {\bf I}_{N_{\rm r}}. \label{eq:RDelayFinal}
\end{equation}
From \eqref{eq:RDelayFinal}, it can be observed that $ \hat{\bf R}^{\hat{\tau}_i}$ has a suitable form for the MUSIC algorithm to be applied. Assuming $N_{\rm r} > |\mathcal{I}_{\tau_i}|$, we can use eigendecomposition of $ \hat{\bf R}^{\hat{\tau}_i} = {\bf E}^{\hat{\tau}_i}_{\rm s} {\bf \Lambda}^{\hat{\tau}_i}_{s} ({\bf E}^{\hat{\tau}_i}_{\rm s})^{\Htran} + {\bf E}^{\hat{\tau}_i}_{\rm n} {\bf \Lambda}^{\hat{\tau}_i}_{\rm n} ({\bf E}^{\hat{\tau}_i}_{\rm n})^{\Htran}$, where the first term corresponds to signal subspace ${\bf \Lambda}^{\hat{\tau}_i}_{\rm s}$ having the significant eigenvalues of $\hat{\bf R}^{\hat{\tau}_i}$ and the second term corresponds to the noise subspace. By using this decomposition, one can write the MUSIC spectrum as
\begin{equation}
    g^{\tau_i}(\theta) = \frac{1}{{\bf a}_{\rm R}^{\Htran} (\theta) {\bf E}^{\hat{\tau}_i}_{\rm n} \left({\bf E}^{\hat{\tau}_i}_{\rm n}\right)^{\Htran} {\bf a}_{\rm R} }. \label{MUSICDelay}
\end{equation}
The angle $\hat{\theta}_{\tau_i}[p]$ corresponding to the $p^{\rm th}$ peak of the MUSIC spectrum $g^{\tau_i}(\theta)$ is the candidate DoA for the $i^{\rm th}$ target.
\begin{remark}
    Multiple targets with identical delay estimates can compromise noise subspace accuracy, reducing MUSIC spectrum peak sharpness. However, leveraging Doppler multiplexing diversity enables more robust DoA estimation despite these subspace limitations. 
\end{remark}
\vspace{-4mm}
\subsection{Doppler Multiplexing}
To eliminate the interference due to targets with different Doppler shifts, filtering on the Doppler axis is proposed to exploit this additional degree of freedom. For this purpose, the filtering along the OFDM symbols for each subcarrier is carried out for the Doppler shift estimate $\hat{f}_i^d$ as
\begin{equation}
   \begin{split}
       &\tilde{\bf y}_{n,l}^{\hat{f}_i} =  \frac{1}{{\sqrt{N_c}}}  \sum_{m = 0}^{M-1} \tilde{\bf y}_{mn,l}  e^{-j2\pi \hat{f}_i^d  m T_{\rm sym}} \\
       &= \sum_{i'=1}^I {\bf a}_{\rm R} (\theta_{i'})  \sum_{m=0}^{M-1}  \bar{\alpha}_{i',mn,l} e^{j2\pi (f_{i'}^d-\hat{f}_i^d) m T_{\rm sym}}  w_{{i'}}^n + \bar{{\boldsymbol \eta}}_{mn,l}. \label{eq:SpectralMF}
   \end{split}
\end{equation}
where $\bar{{\boldsymbol \eta}}_{n,l} \triangleq  \frac{1}{{\sqrt{N_c}}} \sum_{m = 0}^{M-1}  \bar{{\boldsymbol \eta}}_{mn} {\bf x}_{mn}^{*} {{\bf e}_{l}} e^{-j2\pi  \hat{f}_i^d  m T_{\rm sym} }$ and $\bar{\alpha}_{i',mn,l} \triangleq \frac{{\alpha_{i}}}{{\sqrt{N_c}}} {\bf a}^{\Ttran}_{\rm T} (\theta_{i})  {\bf x}_{mn} {\bf x}_{mn}^{*} {{\bf e}_{l}}$. By using a similar reasoning as for \eqref{eq:temporalMF2}, we can simplify \eqref{eq:SpectralMF} as
\begin{equation}
    \tilde{\bf y}_{n,l}^{\hat{f}_i}\approx \sum_{i' \in \mathcal{I}_{f_i}}   \frac{\alpha_{i}}{{\sqrt{N_c}}} {{\bf a}_{\rm R} (\theta_{i'}) } \left[ \sum_{m=0}^{M-1}  \tilde{\bf x}_{mn}[l] \right] w_{{i'}}^m + \bar{{\boldsymbol \eta}}_{n,l}.
\end{equation}
In the next phase, we suppose the intended user has the highest power after Doppler filtering, which determines the potential DoA of the target despite possible inaccuracies due to stronger side-lobe targets. This ambiguity will be resolved by incorporating the DoA with delay processing. To estimate the DoA, the highest peaks in the MUSIC spectrum are our estimation candidates. For this purpose, the auto-correlation function can be approximated with a similar procedure through \eqref{eq:R1Delay}-\eqref{eq:RDelayFinal} as 
\begin{equation}
\begin{split}
    \hat{\bf R}^{\hat{f}_i} &=  \sum_{l=0}^{N_{\rm t}-1} \sum_{n=0}^{N_c-1} \tilde{\bf y}_{n,l}^{\hat{f}_i} \left(\tilde{\bf y}_{n,l}^{\hat{f}_i} \right)^{\Htran} \\
    &{\approx}    \sum_{i' \in \mathcal{I}_{f_i}}  |\alpha_i'|^2 {\bf a}_{\rm R} (\theta_{i'}) {\bf a}_{\rm R}^{\Htran} (\theta_{i'})  \sum_{l=0}^{N_{\rm t}-1} \sum_{n=0}^{N_c-1}  \frac{|\bar{x}_n^l|^2}{{N_c}}   + {N_{\rm t}} N_c {\bf I}_{N_{\rm r}}, \label{eq:RDopplerFinal}
    \end{split}
\end{equation}
where $\bar{x}_n^l \triangleq \sum_{m=0}^{M-1}  \tilde{\bf x}_{mn}[l]$. Assuming $N_{\rm r} > |\mathcal{I}_{f_i}|$, we can use eigendecomposition of $ \hat{\bf R}^{\hat{f}_i} = {\bf E}^{\hat{f}_i}_{\rm s} {\bf \Lambda}^{\hat{f}_i}_{\rm s} ({\bf E}^{\hat{f}_i}_{\rm s})^{\Htran} + {\bf E}^{\hat{f}_i}_{\rm n} {\bf \Lambda}^{\hat{f}_i}_{\rm n} ({\bf E}^{\hat{f}_i}_{\rm n})^{\Htran}$, where the first term corresponds to signal subspace ${\bf \Lambda}^{\hat{f}_i}_{\rm s}$ having significant eigenvalues of $\hat{\bf R}^{\hat{f}_i}$ and the second term corresponds to the noise subspace. By using this decomposition, we can write the MUSIC spectrum as
\begin{equation}
    g^{f_i}(\theta) = \frac{1}{{\bf a}_{\rm R}^{\Htran} (\theta) {\bf E}^{\hat{f}_i}_{\rm n} \left({\bf E}^{\hat{f}_i}_{\rm n}\right)^{\Htran} {\bf a}_{\rm R} }. \label{MUSICDoppler}
\end{equation}
The angle $\hat{\theta}_{f_i}[p]$ corresponding to the $p^{\rm th}$ peak of the MUSIC spectrum $g^{f_i}(\theta)$ is the candidate DoA for the $i^{\rm th}$ target.

\subsection{Fusion of DoA Estimates from Delay/Doppler Domains}

Note that if there are multiple targets in the delay or the Doppler bin used for DoA estimation, there is a significant signal term in the subspace assumed to be the noise subspace. Hence, the term ${\bf a}_{\rm R}^{\Htran} (\theta) {\bf E}^{\hat{\tau}_i}_{\rm n} ({\bf E}^{\hat{\tau}_i}_{\rm n})^{\Htran} {\bf a}_{\rm R}  (\theta)$ or ${\bf a}_{\rm R}^{\Htran} (\theta) {\bf E}^{\hat{f}_i}_{\rm n} ({\bf E}^{\hat{f}_i}_{\rm n})^{\Htran} {\bf a}_{\rm R} (\theta)$ is larger since the subspaces ${\bf E}^{\hat{\tau}_i}$ or ${\bf E}^{\hat{f}_i}$ are in the range space of the target's steering vector. This also means that eigenvalues associated with these subspaces are larger due to strong signal power. Besides, it is not possible to determine which DoA belongs to which target. Hence, we propose the second stage for DoA estimation based on the tentative decision obtained by range/Doppler processing via the MUSIC algorithm. Note that for each range-Doppler pair, we have $4$ angle candidates. For each candidate angle, we perform beamforming and MF to the received signal as

\begin{algorithm}
\caption{$\,$ Summary of the proposed algorithm}\label{alg:Method}
\begin{algorithmic}[1]
\REQUIRE $\bar{\bf y}_{mn}$
\STATE Apply matched filtering in frequency domain as in \eqref{eq:RadarChannels}.

\STATE Construct the data matrix $\bar{{\bf Y}}[k,l]$ as in \eqref{eq:DataMatrix}.

\STATE Obtain the delay/Doppler map for each channel $\boldsymbol{\chi}[k,l]$ via 2D-DFT in \eqref{eq:chiKL}. 

\STATE Perform non-coherent integration in \eqref{eq:NCI} and peak detection for delay/Doppler estimate for $\hat{I}$ targets. \COMMENT{$\hat{I}$ is the number of detected targets.}

\WHILE{$i\leq \hat{I}$}
 \STATE Apply the delay filtering in \eqref{eq:temporalMF} and construct the sample autocorrelation matrix $\hat{\bf R}^{\hat{\tau}_i}$ in \eqref{eq:R1Delay}.
 \STATE Obtain the candidate angles $\hat{\theta}_{\tau_i}[p]$ by using the MUSIC spectrum $g^{\tau_i}(\theta)$ in \eqref{MUSICDelay}.

 \STATE Apply Doppler filtering in \eqref{eq:SpectralMF} and construct the sample autocorrelation matrix $\hat{\bf R}^{\hat{f}_i}$ in \eqref{eq:RDopplerFinal}.
 \STATE Obtain the candidate angles $\hat{\theta}_{f_i}[p]$ by using the MUSIC spectrum $g^{f_i}(\theta)$ in \eqref{MUSICDoppler}.
 \STATE Fuse the candidates by $[\hat{\kappa}_i,\hat{p}] ={\rm argmax}_{\kappa_i,p} \bar{\chi}({\kappa_i},p)$.
 \STATE \textbf{Output}: $\hat{\theta}_i = \hat{\theta}_{\hat{\kappa_i}}[\hat{p}]$
\ENDWHILE

\end{algorithmic}
\end{algorithm}
\begin{equation}
    \bar{{ y}}^{\rm MF}_{mn}({\kappa_i},p) = {\bf a}^{\Htran}_{\rm R}(\hat{\theta}_{\kappa_i}[p]) \bar{{\bf y}}_{mn} ({\bf a}^{\Ttran}_{\rm T}(\hat{\theta}_{\kappa_i}[p]) {\bf x}_{mn} )^*,
\end{equation}
where $\kappa_i \in \{\tau_i,f_i \}$. Then, the data matrix can be constructed as $[\bar{{\bf Y}}^{\rm MF}({\kappa_i},p)]_{mn} = \bar{{ y}}^{\rm MF}_{mn}({\kappa_i},p)$. By using this matrix, one can measure the overall power at the intended delay/Doppler bin for each candidate angle as
\begin{equation}
    \bar{\chi}({\kappa_i},p) = {\bf f}_M[k_i]^{\Ttran} \bar{{\bf Y}}^{\rm MF}({\kappa_i},p) {\bf f}_{N_c}[l_i] .
\end{equation}
After obtaining the power map for candidate DoAs, one can choose the DoA, which gives the maximum power among them as $\hat{\theta}_i = \hat{\theta}_{\hat{\kappa_i}}[\hat{p}]$, where $[\hat{\kappa}_i,\hat{p}] ={\rm argmax}_{\kappa_i,p} \bar{\chi}({\kappa_i},p)$. The proposed algorithm is summarized in Algorithm \ref{alg:Method}. 
\begin{table}[]
\centering
\caption{}
\textsc{System Settings} \break

\begin{tabular}{l|l}
\hline
\textbf{Parameter}                    & \textbf{Value} \\ \hline
Carrier frequency, $f_c$              & $28$ GHz         \\ \hline
Number of subcarriers, $N_c$          & $2048$         \\ \hline
Subcarrier spacing, $\Delta f$        & $30$ kHz    \\ \hline
Number of cyclic prefix, $N_{\rm cp}$ & $144$          \\ \hline
Number of OFDM symbols, $M$           & $256$          \\ \hline
Number of Tx/Rx antennas, $N_t$,$N_r$          & $4$            \\ \hline
\end{tabular} \label{table:SystemParameters} 
\end{table}
\section{Simulation Results}
In this section, we present numerical results demonstrating the performance of the proposed method and compare it with the state-of-the-art algorithms. The parameters related to the system setup are presented in Table~\ref{table:SystemParameters}. Note that the number of Tx/Rx antennas selected is relatively small since the AP is supposed to be employed in a small cell or distributed MIMO system. The $64$ quadrature amplitude modulation (QAM) constellation is used for the OFDM data generation. The target locations are created randomly within a region with radius $150 \, {\rm m}$, and with radial velocity is uniformly selected from the distribution ${\mathcal{U}} \sim [0,50] \, {\rm m/s}$. The noise variance is $-90$\,dBm, and the RCS of targets is assumed to have Gaussian distribution with variance $\sigma_{\rm RCS}^2 = 1$. The two-way path-loss for the target signal is determined via the radar equation as $\lvert \kappa_i \lvert^2= \frac{c^2}{(4\pi)^3f_c^2 d_i^4}$, where $d_i$ is the distance between the target and the BS. To evaluate the performance of the proposed algorithm, it is compared with the benchmark algorithms, namely the ``Data-aided method'' \cite{10634583} and the MUSIC-based ``Sequential method''  \cite{9838368}. 

\begin{figure}
    \centering
    {\includegraphics[width=0.35\textwidth]{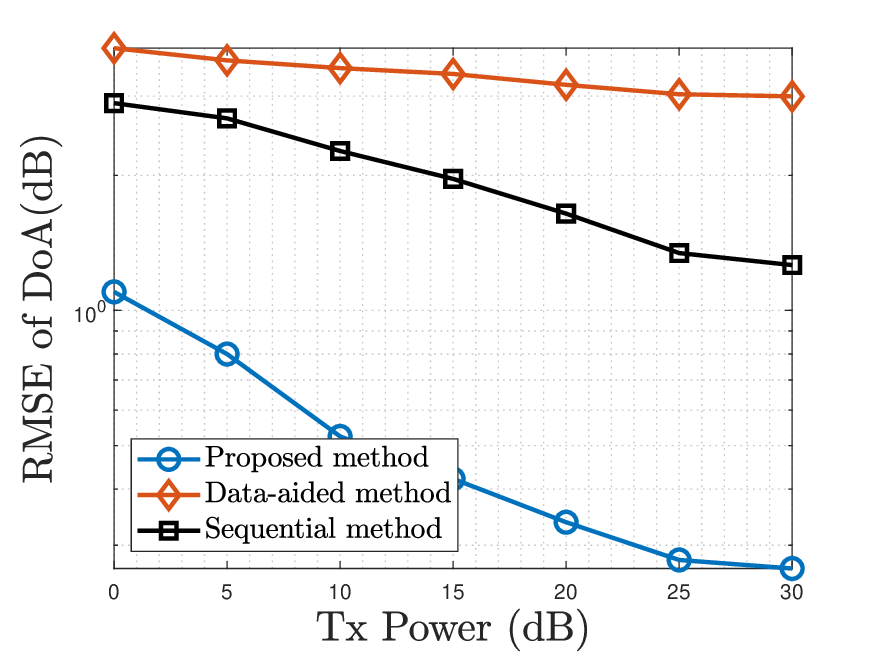}}\hfill
    \caption{Performance vs. the transmit power for $I=3$.}
    \label{RMSETxPower}
\end{figure}

\begin{figure}
    \centering
    {\includegraphics[width=0.35\textwidth]{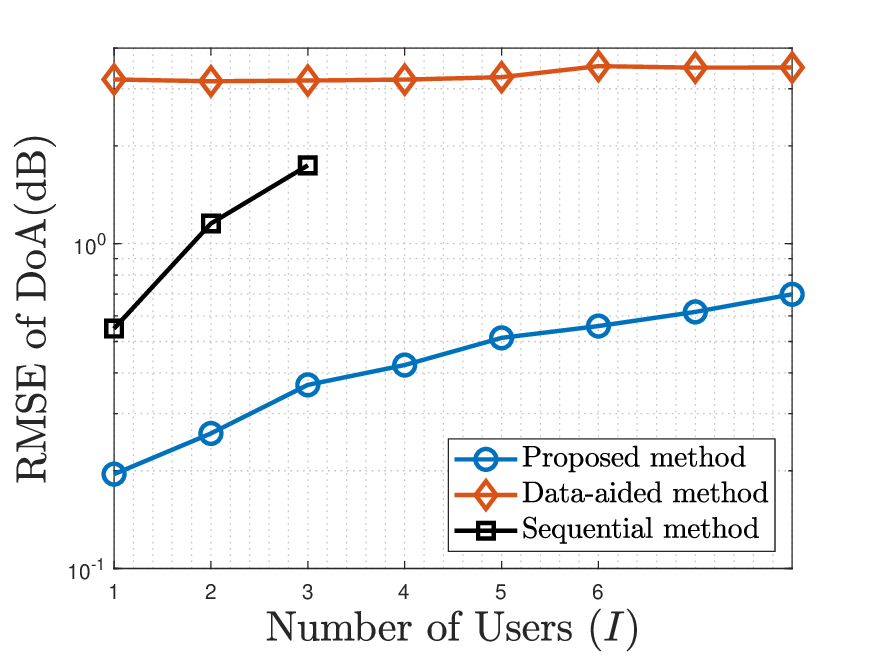}}\hfill
    \caption{Performance vs. the number of users $I$.}
    \label{RMSEUser}
\end{figure}

First, the RMSE performance of the proposed DoA estimation method is evaluated for different total transmit power values at the AP for $I=3$ targets in Fig.~\ref{RMSETxPower}. It is seen that the proposed method outperforms the existing state-of-the-art methods. The proposed algorithm outperforms the data-aided method since the latter relies on the DFT codebook, which has a lower resolution than the super-resolution algorithms. On the other hand, the Sequential method also uses the MUSIC algorithm for the first stage. However, due to strong inter-target interference, this method performs worse. Thanks to the filtering applied at the delay/Doppler multiplexing stage of the proposed algorithm, inter-target interference is suppressed and the signal-to-noise ratio (SNR) is improved thanks to coherent combining over symbols/subcarriers. Hence, the proposed algorithm achieves superior DoA estimation accuracy.

The RMSE performance of the proposed algorithm is investigated with different numbers of targets in Fig.~\ref{RMSEUser}. The results demonstrate that with an AP equipped with $N_{\rm r} =4$ receive antennas, the proposed method exhibits superior DoA estimation performance even for a large number of targets ($I \geq 4$). This enhanced estimation performance cannot be achieved by the conventional sequential estimation algorithm based on the MUSIC method, which faces significant limitations in multi-target scenarios. The results also reveal that the DoA estimation accuracy of the proposed method substantially surpasses its DFT-based data-aided counterpart. Most notably, in scenarios where the number of targets remains below the receive antenna threshold ($N_{\rm r}$), the proposed algorithm demonstrates remarkable superiority over traditional MUSIC-based DoA estimation techniques thanks to its inter-target interference filtering capability.

\begin{figure}
    \centering
    {\includegraphics[width=0.35\textwidth]{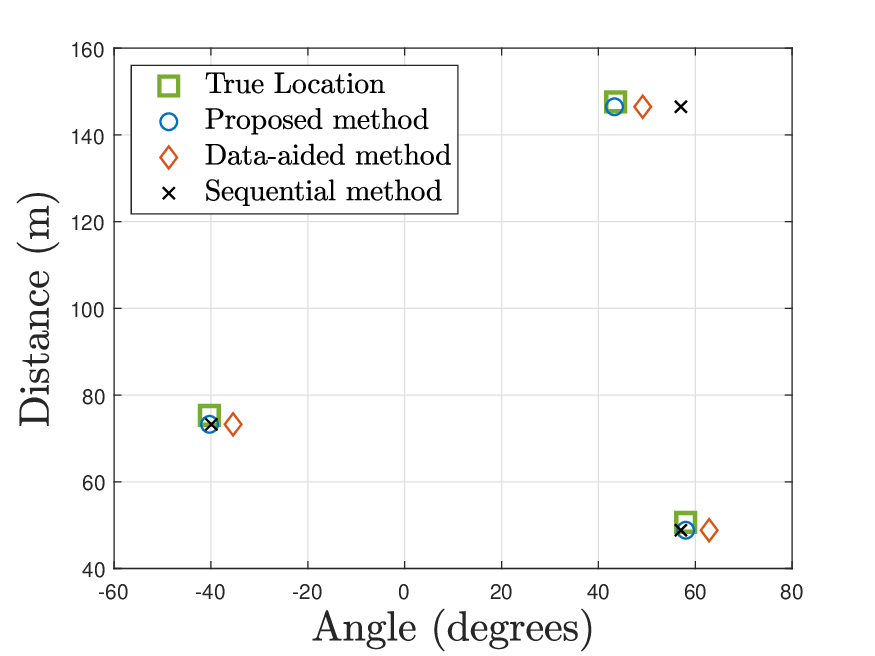}}\hfill
    \caption{Range/angle estimates for $I=3$.}
    \label{3User}
\end{figure}

\begin{figure}
    \centering
    {\includegraphics[width=0.35\textwidth]{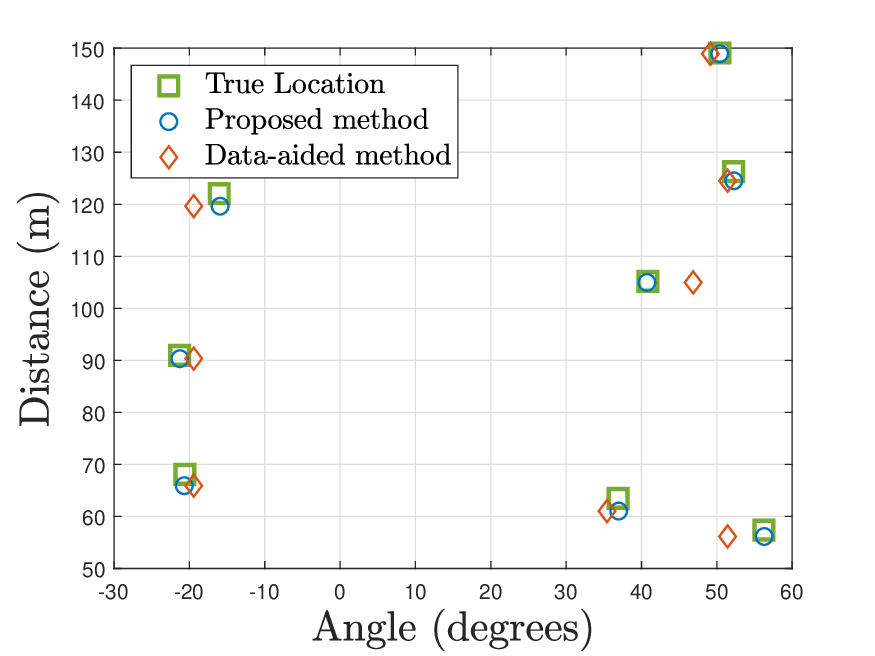}}\hfill
    \caption{Range/angle estimates for $I=8$.}
    \label{8User}
\end{figure}


The estimation performance of the proposed algorithm was evaluated under two different scenarios with $I=3$ and $I=8$ targets, as shown in Fig.~\ref{3User} and Fig.~\ref{8User}. In both cases, the proposed algorithm demonstrated significantly higher accuracy than the benchmark methods. For the $I=3$ scenario Fig.~\ref{3User}, the sequential method could accurately estimate target DoA as the proposed method when the signal-to-interference-plus-noise ratio (SINR) was sufficiently high. In contrast, the DFT-based data-aided method showed notable performance degradation due to its inherent low resolution. Even with a higher number of targets ($I=8$), the proposed algorithm achieved superior performance by leveraging the inter-target interference suppression capability of the MUSIC method as can be seen from Fig.~\ref{8User}.

\section{Conclusions}

This paper addresses the challenges of target sensing in MIMO-radar systems with limited-size antenna arrays at an AP. While existing approaches using DFT-based methods can handle multiple targets but suffer from low angular resolution, and MUSIC-based techniques offer high resolution but are limited by antenna count constraints, our proposed solution bridges this gap. We introduce a novel algorithm that first separates targets in the delay or Doppler domains before performing DoA estimation, enabling effective use of MUSIC even with small antenna arrays. The algorithm incorporates an innovative fusion methodology to mitigate sidelobe interference effects, thereby enhancing DoA estimation accuracy. Through numerical simulations, we demonstrated the superior performance of our approach in achieving high-resolution target sensing in distributed MIMO-ISAC systems.


\bibliographystyle{IEEEtran}
\bibliography{IEEEabrv,refs}

\end{document}